\documentclass[12pt,a4paper]{article}
\usepackage{amssymb}
\usepackage{amsfonts}
\usepackage{amsmath}
\usepackage{float}
\usepackage{latexsym}
\usepackage{cite}
\usepackage{graphicx}
\begin{document}
\title{\vspace{-1cm}\bf Comment on ``Nonperturbative Calculation\\ of Born-Infeld Effects on the Schr\"{o}dinger Spectrum of the Hydrogen Atom''\\
{[Phys. Rev. Lett. 96 (2006) 030402, arXiv:math-ph/0506069]}}
\author{
Mikhail~N.~Smolyakov\\
{\small{\em Skobeltsyn Institute of Nuclear Physics, Lomonosov Moscow
State University,
}}\\
{\small{\em Moscow 119991, Russia}}}
\date{}
\maketitle

\begin{abstract}
It is shown that the solution for the electrostatic potential used in [Phys. Rev. Lett. 96 (2006) 030402, arXiv:math-ph/0506069] is not correct and therefore cannot provide a more accurate spectrum of the hydrogen atom in the Maxwell-Born-Infeld theory than those obtained previously.
\end{abstract}

In Letter \cite{Carley:2006zz}, calculations of the nonrelativistic hydrogen spectrum in the nonlinear Maxwell-Born-Infeld electrodynamics with point charges were presented. As will be demonstrated below, the solution for the electrostatic potential used in \cite{Carley:2006zz} is not correct.

\begin{figure}[ht]
\begin{center}
\includegraphics[width=0.95\textwidth]{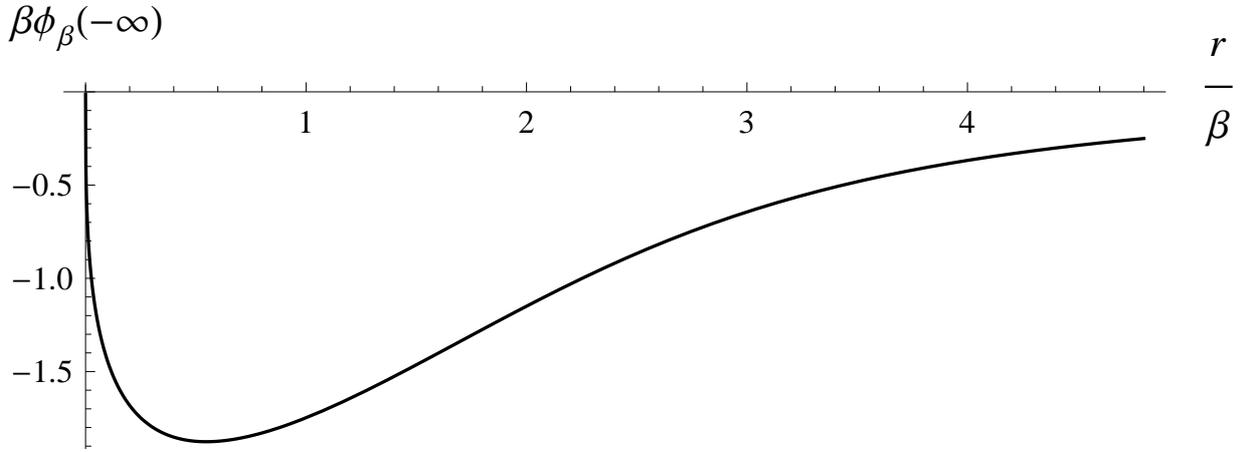}
\caption{Dependence of $\phi_{\beta}(-\infty)$ on $r$.\label{figure1}}
\end{center}
\end{figure}

The analysis in \cite{Carley:2006zz} utilizes the equation
\begin{equation}\label{eqBI}
-\nabla\cdot\frac{\nabla\phi_{\beta}({\bf s})}{\sqrt{1-\beta^{4}|\nabla\phi_{\beta}({\bf s})|^{2}}}=4\pi\left(\delta_{{\bf s}_{p}}({\bf s})-\delta_{{\bf s}_{e}}({\bf s})\right)
\end{equation}
with the standard asymptotic condition $\phi_{\beta}({\bf s})\to 0$ for $|{\bf s}|\to\infty$. It was noted in \cite{Carley:2006zz} that since $\nabla\times{\cal F}_{\mathrm{BI}}({\bf D}_{\mathrm{C}}({\bf s}))\neq 0$, where
${\cal F}_{\mathrm{BI}}({\bf Z})=\frac{{\bf Z}}{\sqrt{1+\beta^{4}{\bf Z}^{2}}}$ and ${\bf D}_{\mathrm{C}}({\bf s})=\frac{{\bf s}-{\bf s}_{p}}{|{\bf s}-{\bf s}_{p}|^{3}}-\frac{{\bf s}-{\bf s}_{e}}{|{\bf s}-{\bf s}_{e}|^{3}}$, almost everywhere, a correct solution of Eq.~(\ref{eqBI}) is
$\nabla\phi_{\beta}({\bf s})=-{\cal F}_{\mathrm{BI}}({\bf D}_{\mathrm{C}}({\bf s})+\nabla\times{\bf G}({\bf s}))$ (to the best of my knowledge, for the first time this important observation was made by Prof.~Kiessling in \cite{Kiessling:2003yg}). However, since $\nabla\times{\cal F}_{\mathrm{BI}}({\bf D}_{\mathrm{C}}({\bf s}))=0$ on the straight line through the point charges, it was assumed that $\nabla\times{\bf G}({\bf s})=0$ for all ${\bf s}$ on this line. In such a case, the electrostatic potential $\phi_{\beta}({\bf s})$ on the straight line through the point charges was defined in \cite{Carley:2006zz} trough the line integral
\begin{equation}\label{def1}
\phi_{\beta}({\bf s})=\int_{{\bf s}}^{\infty}{\cal F}_{\mathrm{BI}}({\bf D}_{\mathrm{C}}({\bf s'}))\,d{\bf s}'.
\end{equation}
Let ${\bf s}\to s$ is the coordinate on the line of integration, ${\bf s}_{p}\to 0$, and ${\bf s}_{e}\to r>0$ is the electron coordinate on this line. It is clear that by definition $\phi_{\beta}(+\infty)=0$. However, it is not so for $\phi_{\beta}(-\infty)$. Indeed, using $s'=xr$ in (\ref{def1}), $\phi_{\beta}(-\infty)$ can be represented as
\begin{align*}
\phi_{\beta}(-\infty)=\frac{2r}{\beta^{2}}\Biggl(\int\limits_{1}^{\infty}\frac{(1-2x)\,dx}{\sqrt{\left(r/\beta\right)^{4}x^{4}(x-1)^{4}
+\left(1-2x\right)^{2}}}\\+\int\limits_{1/2}^{1}\frac{(2x^{2}-2x+1)\,dx}{\sqrt{\left(r/\beta\right)^{4}x^{4}(x-1)^{4}
+\left(2x^{2}-2x+1\right)^{2}}}\Biggr).
\end{align*}
These integrals can be easily evaluated numerically, and it turns out that $\phi_{\beta}(-\infty)\neq 0$ and $\phi_{\beta}(-\infty)$ depends on $r$, see Fig.~\ref{figure1}. Thus, the asymptotic condition for (\ref{def1}) is not satisfied even on the straight line through the point charges.

\begin{figure}[ht]
\begin{center}
\includegraphics[width=0.95\textwidth]{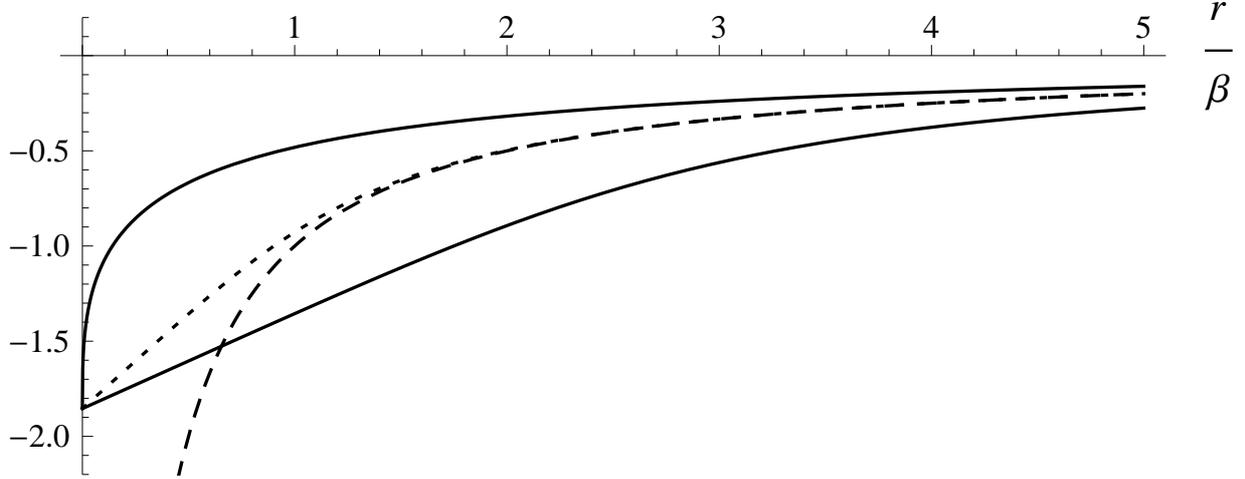}
\caption{Effective potentials of the Schr\"{o}dinger equation: $-(\beta\phi_{\beta}(r)+C/4)$ (upper solid line), $-(\beta\tilde\phi_{\beta}(r)+C/4)$ (lower solid line), $-\beta/r$ (dashed line), and the standard Born-Infeld single particle solution (dotted line). Here $C=B\left(1/4,1/4\right)$ is the beta function. Lower solid line and dashed line reproduce those in Fig.~1 of \cite{Carley:2006zz}.\label{figure2}}
\end{center}
\end{figure}

Furthermore, formula (4) of \cite{Carley:2006zz} was taken from \cite{Kiessling:2003yg}, where it was obtained on the same straight line through the point charges not for definition (\ref{def1}), but for a different definition of $\phi_{\beta}({\bf s})$, namely, for (in the notations of \cite{Carley:2006zz})
\begin{equation}\label{def2}
\tilde\phi_{\beta}({\bf s})=-\int_{{\bf s}_{e}/2}^{{\bf s}}{\cal F}_{\mathrm{BI}}({\bf D}_{\mathrm{C}}({\bf s'}))\,d{\bf s}'.
\end{equation}
It is not difficult to check that $\tilde\phi_{\beta}(\pm\infty)=\mp\phi_{\beta}(-\infty)/2\neq 0$, which means that the asymptotic condition for (\ref{def2}) is not satisfied on the straight line through the point charges as well. As a consequence, definitions (\ref{def1}) and (\ref{def2}) are not equivalent. In particular, $\tilde\phi_{\beta}(r)\not\equiv\phi_{\beta}(r)$. The functions $\phi_{\beta}(r)$ and $\tilde\phi_{\beta}(r)$ can be represented as
\begin{align*}
&\phi_{\beta}(r)=\frac{r}{\beta^{2}}\int\limits_{1}^{\infty}\frac{(1-2x)\,dx}{\sqrt{\left(r/\beta\right)^{4}x^{4}(x-1)^{4}+\left(1-2x\right)^{2}}},\\
&\tilde\phi_{\beta}(r)=-\frac{r}{\beta^{2}}\int\limits_{1/2}^{1}\frac{(2x^{2}-2x+1)\,dx}{\sqrt{\left(r/\beta\right)^{4}x^{4}(x-1)^{4}
+\left(2x^{2}-2x+1\right)^{2}}},
\end{align*}
and it is easy to see that $\phi_{\beta}(r)-\tilde\phi_{\beta}(r)=\phi_{\beta}(-\infty)/2$. Note that the difference between $\phi_{\beta}(r)$ and $\tilde\phi_{\beta}(r)$ is even greater than the difference between $\phi_{\beta}(r)$ (or $\tilde\phi_{\beta}(r)$) and the standard Born-Infeld single particle solution (the latter is used in the test particle approach, which is incorrect as was rightly noted in \cite{Carley:2006zz}), see Fig.~\ref{figure2}.

Thus, for a physically relevant solution of Eq.~(\ref{eqBI}) (i.e., satisfying the asymptotic condition $\phi_{\beta}({\bf s})\to 0$ for $|{\bf s}|\to\infty$) $\nabla\times{\bf G}({\bf s})\neq 0$ on the straight line through the point charges as well. The large difference between $\phi_{\beta}(r)$ and $\tilde\phi_{\beta}(r)$ admits that contribution of the term $\nabla\times{\bf G}({\bf s})$ to the actual solution can be of the order of $\phi_{\beta}(r)$ and $\tilde\phi_{\beta}(r)$ themselves. In principle, it is possible that the actual solution provides the effective potential which is closer to the standard Born-Infeld single particle solution than to $-(\phi_{\beta}(r)+C/(4\beta))$ or $-(\tilde\phi_{\beta}(r)+C/(4\beta))$. However, the term $\nabla\times{\bf G}({\bf s})$ has not been calculated yet, and its real impact on the solution for the electrostatic potential is not clear.

\end{document}